\documentclass{article}

\usepackage[english]{babel}
\usepackage[T1]{fontenc}
\usepackage[utf8]{inputenc}
\usepackage{microtype}

\usepackage[letterpaper,top=2cm,bottom=2cm,left=3cm,right=3cm,marginparwidth=1.75cm]{geometry}

\usepackage{graphicx}
\usepackage{booktabs}
\usepackage{multirow}
\usepackage{array}
\usepackage{tabularx}
\usepackage{threeparttable}
\usepackage{subcaption}
\usepackage{adjustbox}
\usepackage{framed}

\usepackage{amsmath}
\usepackage{siunitx}
\usepackage{arydshln}
\usepackage{xspace}

\usepackage{enumitem}

\usepackage{url}
\usepackage[colorlinks=true, linkcolor=blue, citecolor=blue, urlcolor=blue]{hyperref}

\addto\captionsenglish{%
}
\addto\extrasenglish{%
}

\usepackage[numbers,sort&compress]{natbib}
\renewcommand{\cite}{\citep}

\usepackage{listings}
\usepackage{setspace}

\usepackage{soul,color}
\soulregister\cite7
\soulregister\ref7
\soulregister\pageref7
\setstcolor{red}
\setul{}{.2ex}

\newcommand{\politifactpantsfire}{\texttt{Pants On Fire}\xspace}

\newcommand{\politifactfalse}{\texttt{False}\xspace}
\newcommand{\politifactmostlyfalse}{\texttt{Mostly False}\xspace}
\newcommand{\politifacthalftrue}{\texttt{Half True}\xspace}
\newcommand{\politifactmostlytrue}{\texttt{Mostly True}\xspace}
\newcommand{\politifacttrue}{\texttt{True}\xspace}

\newcommand{\correct}{\textit{correct}\xspace}
\newcommand{\overestimation}{\textit{overestimation}\xspace}
\newcommand{\underestimation}{\textit{underestimation}\xspace}

\newcommand{\standard}{\textit{Standard}\xspace}
\newcommand{\summary}{\textit{Summary}\xspace}

\begin{document}

\title{Efficiency and Effectiveness of LLM-Based Summarization of Evidence in Crowdsourced Fact-Checking}

\author{
Kevin Roitero\textsuperscript{1}, 
Dustin Wright\textsuperscript{2}, 
Michael Soprano\textsuperscript{1}, \\
Isabelle Augenstein\textsuperscript{2}, 
Stefano Mizzaro\textsuperscript{1} \\
\textsuperscript{1}Department of Computer Science, University of Udine, Italy \\
\textsuperscript{2}Department of Computer Science, University of Copenhagen, Denmark \\
\texttt{kevin.roitero@uniud.it}, 
\texttt{dw@di.ku.dk}, 
\texttt{michael.soprano@uniud.it}, \\
\texttt{augenstein@di.ku.dk}, 
\texttt{stefano.mizzaro@uniud.it}
}

\date{}

\maketitle

\begin{abstract}

Evaluating the truthfulness of online content is critical for combating misinformation. This study examines the efficiency and effectiveness of crowdsourced truthfulness assessments through a comparative analysis of two approaches: one involving full-length webpages as evidence for each claim, and another using summaries for each evidence document generated with a LLM. Using an A/B testing setting, we engage a diverse pool of participants tasked with evaluating the truthfulness of statements under these conditions. 

Our analysis explores both the quality of assessments and the behavioral patterns of participants. The results reveal that relying on summarized evidence offers comparable accuracy and error metrics to the standard modality while significantly improving efficiency. Workers in the \summary setting complete a significantly higher number of assessments, reducing task duration and costs. Additionally, the \summary modality maximizes internal agreement and maintains consistent reliance on and perceived usefulness of evidence, demonstrating its potential to streamline large-scale truthfulness evaluations.  
 
\end{abstract}


\section{Introduction}

The proliferation of online misinformation has made truthfulness assessment a crucial task in the digital era. Fact-checking efforts, traditionally conducted by experts, have faced challenges in scalability and timeliness as the volume of information continues to grow \cite{augenstein2024factuality}. Recent developments, such as X and Meta scaling down their reliance on paid fact-checkers,\footnote{\url{https://www.nytimes.com/2025/01/07/business/meta-community-notes-x.html}.} further highlight the urgency of scalable and efficient truthfulness assessment methods \cite{borenstein2025communitynotesreplaceprofessional}. As social media platforms reduce investments in professional fact-checking, alternative approaches become increasingly vital for addressing the growing volume of misinformation. To address this, crowdsourcing has emerged as a promising alternative, leveraging the collective efforts of non-expert workers to evaluate truthfulness efficiently \cite{roitero2020crowd, martel2024crowds}. 
However, the success of crowdsourced truthfulness assessments relies heavily on the format, presentation, and quality of evidence provided to participants. The choice of evidence itself, along with how it is presented, directly influences workers' ability to make accurate and reliable judgments.

Recent advancements in Large Language Models (LLMs) have opened new possibilities for improving the efficiency of truthfulness assessments by generating summarized evidence from lengthy documents. These models can distill essential information into concise formats, potentially reducing cognitive load and speeding up decision-making for crowd workers. Summarization also allows workers to process more evidence in less time, which is particularly valuable in large-scale annotation tasks. However, the adoption of LLM-generated summaries introduces its own set of challenges, including the risk of omitting critical details, introducing biases, or oversimplifying nuanced arguments. 
While summarized evidence could offer advantages in terms of efficiency and scalability, it may also affect the quality of truthfulness assessments. Workers might rely too heavily on the condensed summaries, overlooking important context available in full-length documents. Additionally, inconsistencies or factual errors introduced during the summarization process could inadvertently mislead workers, potentially impacting the reliability of their assessments. These trade-offs necessitate a thorough investigation into both the benefits and limitations of using LLM-generated summaries in crowdsourced truthfulness evaluations.

In this work, we focus on comparing two approaches: a \standard modality, in which participants evaluate full-length webpages, and a \summary modality, in which participants assess summarized evidence generated by a state-of-the-art LLM, \texttt{Meta-Llama-3-8B\-Instruct}~\cite{dubey2024llama}. The summarized format aims to reduce cognitive load and improve worker efficiency by condensing essential information while preserving factual accuracy. This presentation allows participants to process more information in less time, increasing overall throughput. Such improvements not only offer productivity gains but also translate into significant cost savings for large-scale annotation tasks, making summarization particularly valuable. However, the effectiveness and trade-offs of these two modalities remain an open question, which we address through a detailed empirical comparison using an A/B testing framework. This experimental design allows us to systematically compare the two modalities under controlled conditions, providing insights into how evidence presentation affects truthfulness evaluations. To guide our analysis, we address the following Research Questions (RQs):
\begin{enumerate}[label=\textbf{RQ\arabic*}:, left=0em]
    \item \textbf{Effectiveness.} Does the use of summarized evidence achieve comparable accuracy and error metrics to the standard approach of collecting judgments with full-length webpages?

    \item \textbf{Efficiency.} How does the use of summarized evidence impact the time required for participants to complete truthfulness assessments compared to full-length evidence? To what extent can summarization reduce the costs associated with large-scale truthfulness assessment tasks?
        
    \item \textbf{Behavioral Insights.} How does the presentation of evidence influence participants’ behavioral patterns, such as reliance on evidence?
\end{enumerate}

The remainder of this paper is structured as follows:  
Section~\ref{sect:related_work} reviews the background and related work. Section~\ref{sect:methodology} describes our experimental design, including the preparation of data, generation of summarized evidence, the structure of crowdsourcing tasks, and provides demographic statistics. The results are then presented across three key dimensions: effectiveness (Section~\ref{sect:rq-effectiveness}), efficiency (Section~\ref{sect:rq-efficiency}), and worker behavior (Section~\ref{sect:rq-behavior}). Finally, Section~\ref{sect:conclusion} discusses the broader implications of our findings, acknowledges the limitations of the study, and suggests directions for future work.

\section{Background and Related Work}\label{sect:related_work}

Multiple studies have explored the application of crowdsourcing for misinformation detection, demonstrating its potential as a scalable alternative to expert fact-checking \cite{doi:10.1126/science.aao2998,doi:10.1073/pnas.1806781116}. \citet{la2020crowdsourcing} highlighted the influence of worker bias in assessing statements, an issue further explored by \citet{roitero2020crowd}, who observed similar agreement levels across various truthfulness scales. Expanding on this, \citet{SOPRANO2021102710} and \citet{liu2024exploring} introduced multidimensional truthfulness scales, showing that crowd judgments are reliable across multiple truthfulness dimensions, each capturing different aspects of truthfulness. The comparative performance of crowd workers and experts has also been examined. For example, \citet{zhao2023variety} found that while the crowd performed on par with experts in terms of accuracy, objectivity, and clarity, their assessments often depended on existing professional knowledge. Similarly, \citet{saeed2022crowdsourced} analyzed social network data and demonstrated that crowd evaluations relied on different evidence sources compared to experts, with better scalability and efficiency. \citet{allen2022birds} further highlighted the scalability of crowd-based methods, noting that political alignment influenced crowd judgments. Additionally, linguistic \cite{10.1145/3331184.3331248} and psychological \cite{pennycook2021psychology} peculiarities in misinformation were explored, emphasizing the diverse factors affecting the quality of crowdsourced truthfulness evaluations.

While crowdsourcing holds significant potential, it is not without challenges \cite{10.1145/3159652.3159654,10.1145/3290605.3300637}. \citet{draws2022bias} identified a general tendency among workers to overestimate truthfulness, often driven by systematic cognitive biases \cite{SOPRANO2024103672,draws2021checklist}. Bias-aware aggregation methods have been proposed to mitigate these distortions, such as the confirmation bias addressed by \citet{gemalmaz2021accounting}. Additionally, \citet{10.1145/3397271.3401408} presented FactCatch, a human-in-the-loop system designed to minimize effort while guiding users in fact-checking tasks. Further analyses have explored how biases can be identified and corrected to improve the quality of crowdsourced assessments. For example, \citet{10.1145/3289600.3291035} investigated task abandonment patterns, while \citet{10.1145/3586998} proposed methods to enhance labeling reliability by leveraging correlations between labels and neighboring instances. These efforts underscore the importance of addressing cognitive and procedural challenges to maximize the reliability of crowdsourced truthfulness assessments.

Although this study focuses on crowdsourcing, advances in automatic and hybrid fact-checking methods have contributed significantly to the field. Fully automated techniques, mostly based on machine learning models, have been widely explored for misinformation detection \cite{augenstein2024factuality, dmonte2025claimverificationagelarge, https://doi.org/10.48550/arxiv.2102.04458, doi:10.1080/24751839.2020.1847379, 8862770, HU2022133, atanasova-etal-2022-fact}. These methods have been evaluated using public datasets and benchmarks, including those provided by initiatives like the CLEF CheckThat! Lab \cite{10.1007/978-3-031-13643-6_29, 10.1007/978-3-030-72240-1_75} as well as other projects such as MultiFC \cite{augenstein-etal-2019-multifc}, FEVER \cite{thorne-etal-2018-fever, thorne-etal-2019-fever2}, AVeriTeC \cite{schlichtkrull-etal-2024-automated} and Factcheck-Bench \cite{wang-etal-2024-factcheck}. Comparative analyses also have been developed to study variability in model performance across datasets \cite{8843612,HU2022133,riedel2017simple,10.1145/3626772.3661361}. Recent research has also focused on generating human-readable explanations \cite{atanasova-etal-2020-generating-fact, 10.1145/3546917, 10.1145/3539618.3591879, 10.1145/3437963.3441828} and incorporating complex interactions within machine learning pipelines \cite{10.1145/3686962}. Hybrid approaches that integrate human input with machine learning have gained attention as well \cite{9923043,9648388,10.1145/3546916}. Lastly, there is research on how fact checkers use tools in practice \cite{Warren_2025}.

In this paper, we leverage automatic signals generated by a state-of-the-art LLM, \texttt{Meta-Llama-3-8B\-Instruct} \cite{dubey2024llama}, to enhance the efficiency and scalability of a crowd-based approach for truthfulness assessment. By combining LLM-generated summaries with crowdsourced evaluations, we aim to investigate the interplay between automated summarization and human judgment in achieving reliable and cost-effective misinformation detection.

\section{Methodology}
\label{sect:methodology}

In this section, we detail the methodology employed in our experiment. To promote transparency and reproducibility, all data collected during the experiment are made publicly available to the research community.\footnote{\url{https://doi.org/10.17605/OSF.IO/BNFXM}.}

\subsection{Data}

This study uses the same dataset and adopts a similar experimental design to that reported by \citet{BARBERA2024103792}, with modifications tailored to meet the specific objectives and requirements of our research. To ensure that this paper is self-contained and understandable independently, we report both the details of the dataset and the crowdsourcing task employed.
The dataset consists of a subset of statements sourced from the PolitiFact website,\footnote{\url{https://www.politifact.com/}} as in previous research \cite{SOPRANO2021102710, draws2022bias, BARBERA2024103792}. PolitiFact has been an active platform for fact-checking statements made by U.S. politicians, political organizations, public figures, and social media posts since 2007, accumulating over 24,000 fact-checks with regular updates.

PolitiFact statements are categorized by expert judges using a six-level truthfulness scale, consisting of the following levels: \politifactpantsfire, \politifactfalse, \politifactmostlyfalse, \politifacthalftrue, \politifactmostlytrue, and \politifacttrue. In this study, we adopt the same six-level scale to maintain consistency and enable direct comparisons between expert fact-checkers and crowdsourced assessments. As in previous work, we select a subset of 120 statements (20 for each ground truth category) from PolitiFact. Additionally, we ensure that each statement is supported by at least 10 webpages retrieved as evidence using the Bing Web Search API.\footnote{\url{https://www.microsoft.com/en-us/bing/apis/bing-web-search-api}} These statements form the basis of the crowdsourcing experiments described in the subsequent sections.

\subsection{Evidence Summarization}

\begin{table}[tb]
    \centering
\caption{Example of an LLM-generated summary given the query `Opposition to having a fully elected Chicago Board of Education is in the ``super minority.'''.}
\label{tab:summary}
\small
\begin{tabular}{p{5cm}@{\quad}p{9cm}}
\toprule
\textbf{Evidence} & \textbf{Summarized Evidence} \\
\midrule
For the first time in Chicago's history, voters would get a say in who runs the city's school board under two competing proposals now before the state Legislature. One calls for a fully elected Board of Education, and the other for a ``hybrid'' model splitting
the school board into some elected members with the majority still appointed
by the mayor.  [...]
& 
- The document discusses the possibility of having a fully elected Chicago Board of Education, with a focus on the overwhelming support for this idea among Chicagoans, as shown through various polls and referendums.

- The current system of political appointment by the mayor has been in place for 150 years, but there is no consensus among researchers on which form of governance fosters better student performance or fiscal management.

- Over 90\% of school districts nationwide have elected boards, and overwhelming majorities of Chicagoans have long favored a switch to an elected board. [...]
\\
\bottomrule
\end{tabular}
\end{table}

We process the evidence related to each statement to create two distinct versions for evaluation: the original, unaltered full-length web page and a more concise summarized version, obtained by leveraging a LLM. To achieve the summarization version, we employ the capabilities of the \texttt{Meta-Llama-3-8B-Instruct} model, a state-of-the-art model known for its robust performance in natural language understanding and generation tasks \cite{dubey2024llama}.\footnote{\url{https://huggingface.co/meta-llama/Meta-Llama-3-8B-Instruct}}

The process begins with a carefully crafted prompt provided to the model. This prompt is specifically designed to guide the model in extracting and summarizing the essential factual elements from the content of a webpage, ensuring that the generated summary maintains consistency with the original information in terms of factuality and stance. After several rounds of trial and error, the exact wording of the prompt we use is given in Figure~\ref{fig:llama_prompt}.

\begin{figure}[tb]
    \centering
    \fbox{
        \parbox{0.9\linewidth}{
            \textbf{System Prompt:} \\
            You are a helpful, respectful, and honest assistant. Your job is to provide a summary of the provided documents based on a user query. Your summaries should be as accurate as possible and should not include any harmful, unethical, racist, sexist, toxic, dangerous, or illegal content. Please ensure that your responses are socially unbiased and positive in nature.
            
            \vspace{0.5em}
            \textbf{User Prompt:} \\
            Here is a user query: \texttt{\{statement\}}. Please do your best to provide a concise list of points which summarize the following document. Please focus on points related to the query. Please output your response as a JSON object with a field \texttt{"summary"} that contains the list of strings. Please first describe what the document is. Then please output five or more strings summarizing the contents of the document. Here is the document: \texttt{\{document content\}}.
        }
    }
    \caption{Evidence summarization prompt.}
    \label{fig:llama_prompt}
\end{figure}
Using this prompt, the model generates a summary that distills the webpage's content into a shorter form, focusing on the most critical information needed to perform the fact-checking task. An example of the summarized evidence is shown in Table~\ref{tab:summary}.

\subsection{Crowdsourcing Tasks}

We adopt the crowdsourcing task as detailed in \citet{BARBERA2024103792}, modifying it to collect responses for our study while preserving the core structure of the original design. This design, which was originally developed and validated in earlier research \cite{BARBERA2024103792, SOPRANO2021102710, draws2022bias}, allows us to collect truthfulness assessments at scale using crowdsourcing.

We employ a task design featuring sequential data collection, where participants evaluate statements one at a time for truthfulness. We recruit participants through the Prolific platform and apply a filter to include only workers from the U.S. Compensation reflects fair market practices and is based on careful estimations of task completion times, ensuring that payments are both ethical and practical. For the \standard modality, we pay each worker \textsterling{}2, while for the \summary modality, we set compensation at \textsterling{}1.80. Each worker evaluates a total of 8 statements (one for each ground truth category and two additional gold-standard questions), and 5 distinct workers assess each statement. We recruit 100 workers for each modality, resulting in a total cost of approximately \textsterling{}500, including Prolific's fees.

After filling a demographic questionnaire, the participants perform our task using a structured interface where they first encounter the statement to be assessed, followed by a set of evidence. Each worker sees the title of the web page along with a snippet, similar to the presentation layout of a search engine. Then, depending on the task layout, if the worker clicks on the link they see either the full webpage or its summary. Subsequent to evidence review and selection, participants assess the statement. We ask workers to first evaluate the overall truthfulness of the statement using the same six-level scale as the one used by experts. Then, we ask the workers to answer two questions regarding how useful they found the evidence in determining the truthfulness of the statement: ``How useful was the evidence?'' and ``Did you use the evidence to fact-check the statement, or did you have a clear opinion before reviewing the evidence?''. For both questions, we collect responses on a five-level Likert scale. The scale for the first question includes the following options: ``Not Useful At All'', ``Slightly Useful'', ``Moderately Useful'', ``Very Useful'', and ``Extremely Useful''. The scale for the second question includes: ``I did not consider the evidence at all before forming my opinion'', ``I considered the evidence slightly before forming my opinion'', ``I considered the evidence as much as my initial opinion before forming my final opinion'', ``I considered the evidence more than my initial opinion before forming my final opinion'', and ``I fully based my opinion on the evidence after reviewing it''.
To ensure high-quality assessments from workers, we implement the following checks: a minimum time requirement of 3 seconds per statement and correct evaluations of two gold-standard questions, one clearly true and the other clearly false. 

\subsection{Descriptive Statistics}\label{sect:descriptive}

\begin{table}[t]
    \centering
    \caption{Demographic distribution across modalities.}
    \label{tab:demographics}
    \begin{tabular}{lrr}
    \toprule
    \textbf{Consideration} & \textbf{\standard (\%)} & \textbf{\summary (\%)} \\
    \midrule
        Democrat & 40 & 45 \\
        Independent & 17 & 23 \\
        Republican & 40 & 31 \\
        Something Else & 3 & 1 \\
        \midrule
    \textbf{Political Views} & \textbf{\standard (\%)} & \textbf{\summary (\%)} \\
    \midrule
        Conservative & 25 & 23 \\
        Liberal & 24 & 26 \\
        Moderate & 23 & 25 \\
        Very Conservative & 11 & 8 \\
        Very Liberal & 17 & 18 \\
    \bottomrule
\end{tabular}
\end{table}

The answers to the demographic questionnaire enable us to assess the demographics of participants across the two experimental conditions. Table~\ref{tab:demographics} presents the distribution of political considerations and views for each condition. The results indicate a balanced representation across both modalities, with comparable percentages for both political affiliation and ideological orientation. 

We also measure abandonment \cite{10.1145/3289600.3291035} to evaluate participant engagement. Initial abandonment rates (49\% for \standard, 38\% for \summary) and mid-task abandonment rates (10\% for \standard, 13\% for \summary) are comparable, with no statistically significant differences observed.

\section{RQ1: Effectiveness}\label{sect:rq-effectiveness}

\subsection{Agreement with Experts}

\begin{figure*}[tb]
    \centering
    \begin{tabular}{c}
    \includegraphics[width=\linewidth]{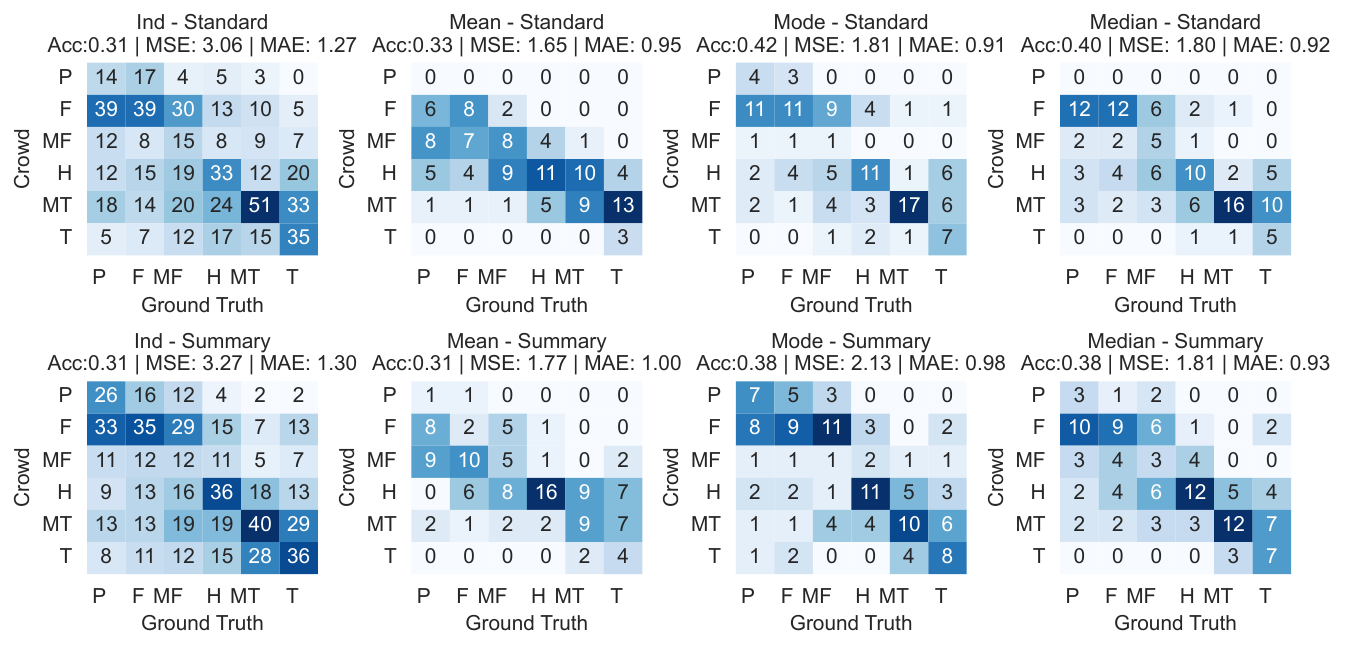}
    \end{tabular}
    \caption{Agreement between workers and experts on individual and aggregated assessments.}
\label{fig:external}
\end{figure*}

We inspect the agreement between workers and experts, shown in Figure~\ref{fig:external}, where we report results for both individual and aggregated assessments. In all figures, the six PolitiFact levels are abbreviated with their initials. We discuss both the individual and the aggregated scores for the \standard and \summary modalities. In terms of individual scores, in the standard modality, we observe an accuracy of 0.31, a MSE of 3.06, and a MAE of 1.27. The summary modality reports a similar accuracy (0.31) but exhibits a slightly higher MSE (3.27) and MAE (1.30). This similarity in the metric scores suggests a comparable quality level between the standard and summary modalities, indicating that the summarized content does not impact workers' effectiveness. This finding is nontrivial, given that the content presented in the summary version of the task is substantially condensed.

We further examined whether the observed differences in MAE and MSE between the modalities are statistically significant. Using bootstrapped confidence intervals\footnote{\url{https://docs.scipy.org/doc/scipy/reference/generated/scipy.stats.bootstrap.html}.} with 10,000 resamples, we estimated the range of plausible values for the difference in these metrics. For MAE, the mean difference was \(-0.028\) with a 95\% confidence interval ranging from \(-0.167\) to \(0.112\). Similarly, for MSE, the mean difference was \(-0.212\) with a 95\% confidence interval spanning from \(-0.767\) to \(0.340\). As both intervals include zero, we conclude that there is no statistically significant difference between the standard and summary modalities in terms of MAE and MSE.

We now turn to the aggregated scores, which are still shown in Figure~\ref{fig:external}. The scores are aggregated using three distinct statistical methods: mean, mode (i.e., majority voting), and median. The mean offers a general view of central tendency and has proven to lead to higher agreement with expert assessments \cite{BARBERA2024103792, SOPRANO2021102710}, while the mode identifies the most frequently occurring outcome, and the median provides an additional robust measure of central tendency that is less sensitive to extreme values.
In the standard modality, mean accuracy is slightly higher (\(0.33\)) compared to the summary modality (\(0.31\)), though the difference is minimal. Similarly, the standard modality shows marginally better error metrics, with MSE (\(1.65\) vs. \(1.77\)) and MAE (\(0.95\) vs. \(1.00\)) indicating lower dispersion and deviation from the ground truth. Mode aggregation reflects a similar trend, with accuracy at \(0.42\) for the standard modality and \(0.38\) for the summary modality, alongside smaller errors (MSE: \(1.81\) vs. \(2.13\), MAE: \(0.91\) vs. \(0.98\)). Median aggregation further supports this finding, with the standard modality showing slightly better performance across metrics (accuracy: \(0.40\) vs. \(0.38\), MSE: \(1.80\) vs. \(2.13\), MAE: \(0.92\) vs. \(0.93\)). While errors in the summary modality are slightly more pronounced, both approaches achieve comparable accuracy levels. Bootstrap analysis confirms no statistically significant differences across accuracy and error measures. Overall, both modalities show similar effectiveness across several metrics, indicating that, despite the reduced volume of information provided to workers in the summary modality, it closely resembles the standard modality in terms of data quality.
To test whether worker demographics influence the effectiveness of truthfulness assessments, we additionally run an Ordinary Least Squares (OLS) regression analysis \cite{freedman2009statistical}. The results show no statistically significant relationship between demographics and accuracy, indicating that these factors do not substantially impact assessment performance.

\subsection{Agreement Among Workers}

 \begin{figure}[tb]
    \centering
    \begin{tabular}{c}
    \includegraphics[width=.65\linewidth]{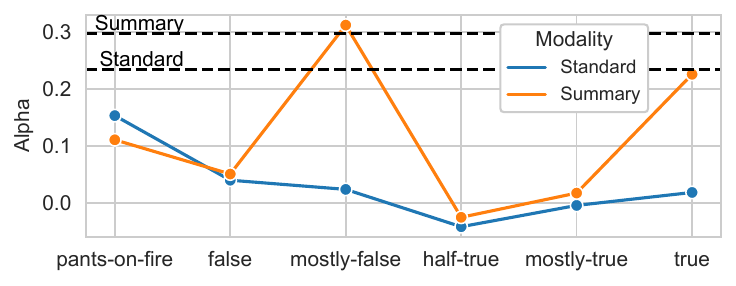}
    \end{tabular}
    \caption{Krippendorff's $\alpha$ score.}
    \label{fig:alpha}
\end{figure}

We employ Krippendorff's $\alpha$ \cite{krippendorff2011computing} to assess the internal agreement among participants in the two task variations. Krippendorff's $\alpha$ is a robust statistical measure used to determine the consistency of ratings provided by different evaluators, and it is a very popular measure for crowdsourcing tasks \cite{Checco_Roitero_Maddalena_Mizzaro_Demartini_2017}.

Figure~\ref{fig:alpha} shows the values of internal agreement across the two task modalities, both for the whole task (dotted lines) as well as for the single ground truth levels of the statements.
For the \standard modality, the $\alpha$ values are generally lower than the \summary modality, reflecting overall less agreement among participants, with the only exception of the \politifactpantsfire truthfulness level. This suggests that when participants are presented with full-length content, there may be greater diversity in how the information is interpreted, possibly due to the complexity and volume of the provided information. Furthermore, the lowest $\alpha$ values appear in the categories in the middle of the scale, indicating that the claims with no extreme truthfulness score are also the most challenging for participants to consistently evaluate.
Conversely, the \summary modality shows higher $\alpha$ values across all categories except \politifactpantsfire and for the overall task. This difference is statistically significant at the $p<0.05$ level, according to the Mann-Whitney U Test \cite{mann1947test}, indicating a higher degree of consistency among workers' ratings. This improvement in internal agreement could be attributed to the condensed nature of the content, which might focus the workers evaluations and reduce both the cognitive load as well as the ambiguity or variability in interpretation that more extensive evidence might introduce. Notably, the alpha values are particularly higher in categories like \politifactmostlyfalse and \politifacttrue where the quality of the summarization may help clarify the key aspects of the statements.

Overall, these findings remark the impact of evidence summarization on the consistency of workers assessments. The higher internal agreement in the summary modality suggests that summarization not only maintains the essential content needed for the truthfulness assessment but may also aid in aligning worker scores, leading to more uniform interpretations and assessments of the truthfulness of statements.

\subsection{Error Analysis}
\begin{table}[tb]
\centering
\caption{Error types across truthfulness levels.}
\label{tab:merged_error_analysis}
\begin{tabular}{llccc}
\toprule
\textbf{Modality} & \textbf{Scope} & \textbf{Correct} & \textbf{Over} & \textbf{Under} \\
\midrule
\standard   & Overall        & 187 (30.6\%) & 237 (38.8\%) & 176 (28.8\%) \\
\summary    & Overall        & 185 (30.2\%) & 232 (37.9\%) & 183 (29.9\%) \\
\midrule
\multirow{6}{*}{\standard} 
    & \politifactpantsfire   & 24 (24.0\%)   & 76 (76.0\%)   & -- \\
    & \politifactfalse           & 39 (37.1\%)   & 44 (41.9\%)   & 35 (21.0\%) \\
    & \politifactmostlyfalse   & 15 (18.3\%)   & 51 (62.2\%)   & 34 (19.5\%) \\
    & \politifacthalftrue      & 33 (48.5\%)   & 41 (30.1\%)   & 26 (20.6\%) \\
    & \politifactmostlytrue    & 51 (50.0\%)   & 15 (14.7\%)   & 34 (33.3\%) \\
    & \politifacttrue            & 35 (35.0\%)   & --            & 65 (65.0\%) \\
\midrule
\multirow{6}{*}{\summary} 
    & \politifactpantsfire   & 26 (26.0\%)   & 74 (74.0\%)   & -- \\
    & \politifactfalse          & 35 (35.7\%)   & 49 (50.0\%)   & 16 (14.3\%) \\
    & \politifactmostlyfalse   & 12 (15.6\%)   & 47 (61.0\%)   & 41 (23.4\%) \\
    & \politifacthalftrue     & 36 (42.9\%)   & 34 (40.5\%)   & 13 (16.6\%) \\
    & \politifactmostlytrue     & 40 (50.6\%)   & 28 (32.0\%)   & 32 (17.4\%) \\
    & \politifacttrue           & 36 (36.0\%)   & --            & 64 (64.0\%) \\
\bottomrule
\end{tabular}
\end{table}

To better understand the patterns of errors in truthfulness assessments, we conducted a failure analysis by categorizing worker assessments into three types: \correct, \overestimation, and \underestimation. The results, shown in the first part of Table~\ref{tab:merged_error_analysis} indicate that the distribution of error types is similar across the two modalities. These comparable distributions suggest that the modality of evidence presentation does not significantly affect the overall patterns of worker errors.

A more fine-grained analysis, shown in the second part of Table~\ref{tab:merged_error_analysis}, highlights the distribution of errors across modalities. 
In the \standard modality, \textit{overestimation} emerges as the most frequent error type for categories such as \politifactmostlyfalse (62.2\%). Conversely, \correct assessments are most common in categories like \politifactmostlytrue (50.0\%) and \politifacthalftrue (48.5\%), suggesting workers find it easier to correctly evaluate statements closer to the middle of the truthfulness spectrum. \textit{Underestimation} occurs less frequently overall but is more pronounced for higher truthfulness categories.
In the \summary modality, \textit{overestimation} remains dominant in the same categories as in the \standard modality. However, \textit{correct} assessments are slightly more evenly distributed, with the highest rates observed for \politifactmostlytrue (50.6\%) and \politifacthalftrue (42.9\%). Overall, these findings suggest that both modalities exhibit similar patterns with respect to error types.
We further analyze the mismatch magnitude for \textit{overestimation} and \textit{underestimation} errors across the \standard and \summary modalities and find similar error patterns. Both modalities exhibit comparable average mismatch values for \textit{overestimation} (1.99) and slightly more negative values for \textit{underestimation} in the \summary modality (-1.73 vs. -1.65). None of these differences are statistically significant, confirming the overall comparability of the two approaches.

\subsection{Robustness to Fewer Judgments}

\begin{figure}[tb]
    \centering
    \begin{tabular}{c}
    \includegraphics[width=.63\linewidth]{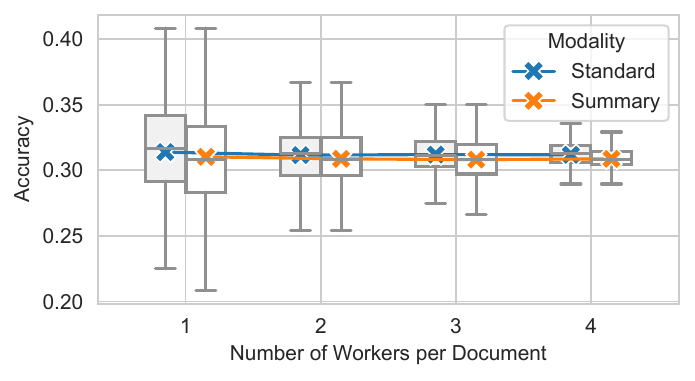}
    \end{tabular}
    \caption{Accuracy across sampling sizes.}
    \label{fig:bootstrap}
\end{figure}

To investigate the robustness of the agreement between worker and expert truthfulness evaluations, we conduct a bootstrap analysis \cite{efron1994introduction}. This method involves resampling the collected assessments with replacement, enabling us to estimate the variability in effectiveness metrics across different sampling sizes. By applying this approach, we aim to understand how the number of workers per document impacts the overall accuracy and consistency of aggregated truthfulness assessments in both task modalities. For each sampling size, ranging from 1 to 4 workers per document (5 would include all workers), we perform 1,000 repetitions during which we compute the individual accuracy of the aggregated assessments compared to expert evaluations. 

The results, shown in Figure~\ref{fig:bootstrap}, reveal a trend in which the accuracy of both modalities slightly improves, although without statistical significance, and the variance decreases as the number of workers per document increases. This pattern is consistent with expectations from crowdsourcing settings, where aggregating multiple assessments tends to mitigate individual biases and errors. Notably, the \summary modality achieves effectiveness levels comparable to the \standard modality across all sampling sizes. These indicate that summarized evidence can provide comparable results to full-length evidence in terms of both accuracy and consistency, and that this equivalence is achieved while potentially reducing the cognitive load on workers and the resources required for task completion.

\section{RQ2: Efficiency}\label{sect:rq-efficiency}

\subsection{Time}
\begin{figure}[tb]
    \centering
    \begin{tabular}{c}
    \includegraphics[width=.63\linewidth]{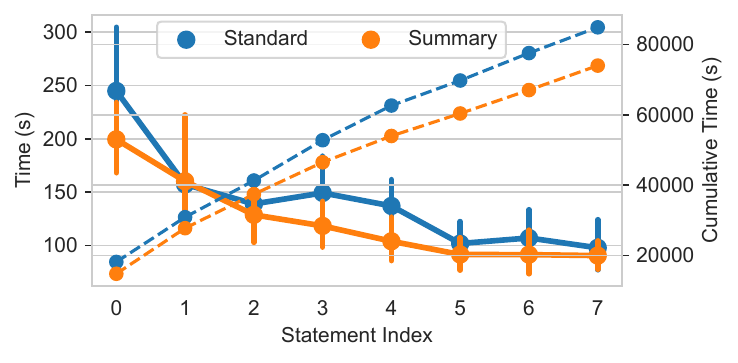}
    \end{tabular}
    \caption{Time elapsed on statements for the two modalities.}
    \label{fig:time}
\end{figure}

In evaluating the practical aspects of crowdsourced truthfulness assessments, time efficiency emerges as a critical metric, as the more time a worker spends on the task, the higher the cost required for their judgment and, consequently, for the overall task.
Figure~\ref{fig:time} shows the time required for workers to provide assessments for both task modalities. As shown by inspecting the two series in the plot, the average time taken to complete assessments consistently decreases as the statement index increases (i.e., as a worker performs the task). This trend suggests that participants become more proficient at navigating and processing the tasks over time. Such a learning effect, also observed in previous studies \cite{BARBERA2024103792, SOPRANO2021102710}, suggests the adaptability of participants and their increasing efficiency in handling the assessments, whether they are dealing with full-length or summarized content.

When comparing the two series we see that the \summary modality consistently shows a reduced average time on each statement index compared to the \standard modality, with the only exception being the second document, and the difference is statistically significant according to the Mann-Whitney U Test \cite{mann1947test} at the  $p<0.05$ level. This supports the observation that time efficiency gained from presenting workers with summarized evidence are real. In practice, the reduced time spent in the \summary modality can lead to enhanced productivity and a lighter cognitive load for participants, which can perform more judgment in the same amount of time and are required to read much less text. Such improvements in efficiency could also translate into cost savings for researchers and practitioners implementing these misinformation assessments at scale, potentially enhancing the economic feasibility of deploying crowdsourced truthfulness evaluations. This can be seen by inspecting the cumulative series in Figure~\ref{fig:time}, which show the potential for cost savings in large-scale truthfulness assessments.

For instance, completing 600 assessments, equivalent to 120 documents, requires approximately 23.67 hours of total work and costing \pounds\xspace{}171.60 at the U.S. minimum wage of \$7.25 per hour (approximately \pounds\xspace{}6.08). In contrast, the \summary modality, reduces the required total work to approximately 20.50 hours, costing \pounds\xspace{}148.63 under the same wage assumption. While these figures are based on the minimum wage, actual costs would scale proportionally if higher wages are offered.  When extended to larger datasets requiring (let us say) 6,000 assessments, the \standard modality incurs costs of \pounds\xspace{}1,716, compared to \pounds\xspace{}1,486 for the \summary modality, resulting in higher savings. 
These findings emphasize the economic advantages of the \summary modality, which maintains comparable performance metrics while significantly reducing worker time and overall costs, making it particularly valuable for large-scale annotation efforts.
 
\subsection{Effectiveness Comparison Under Equal Time Constraints}

To evaluate the effectiveness of evidence summarization under equal time constraints, we compared the effectiveness of the two modalities across metrics. Additionally, we assessed the efficiency of each modality in terms of the number of assessments completed within the same time frame, highlighting the potential advantages of summarization for throughput.
To ensure a fair comparison, we simulate equal time constraints for both modalities by aligning the total time spent on assessments for each document in the dataset. For every document in the \standard modality, we compute the total time workers spend on assessments. Then, we iteratively stack assessments from the corresponding \summary modality until their cumulative time matches or exceeds the total time recorded for the \standard modality. Then, we compute metrics separately for each document in both modalities. We use the \standard modality as the baseline and express the relative efficiency of the \summary modality as a judgment multiplier, calculated as the ratio of completed assessments in the \summary modality to those in the \standard modality. 

\begin{table}[tb]
\centering
\caption{Comparison under equal time constraints.}
\label{tab:equal_time_comparison}
\begin{tabular}{lrr}
\toprule
\textbf{Metric} & \textbf{\standard Value} & \textbf{\summary Value} \\
\midrule
Accuracy         & 0.312                   & 0.308                  \\
MSE              & 3.063                   & 3.275                  \\
MAE              & 1.270                   & 1.298                  \\
Judgment Multiplier & --                    & $+15\%$                  \\
\bottomrule
\end{tabular}
\end{table}

Table~\ref{tab:equal_time_comparison} presents the comparative results. The \standard modality achieves a slightly higher accuracy (\(0.312\)) compared to the \summary modality (\(0.308\)), with minimal differences in error metrics. These findings indicate comparable performance in judgment quality between the two modalities, also considering that no statistical significance is detected.
The primary advantage of the \summary modality thus lies in its efficiency. On average, workers in the \summary setting completed about 15\% more assessments than those in the \standard setting, offering significant efficiency gains and cost savings, as detailed earlier.

\begin{figure*}[tb]
    \centering
    \includegraphics[width=\linewidth]{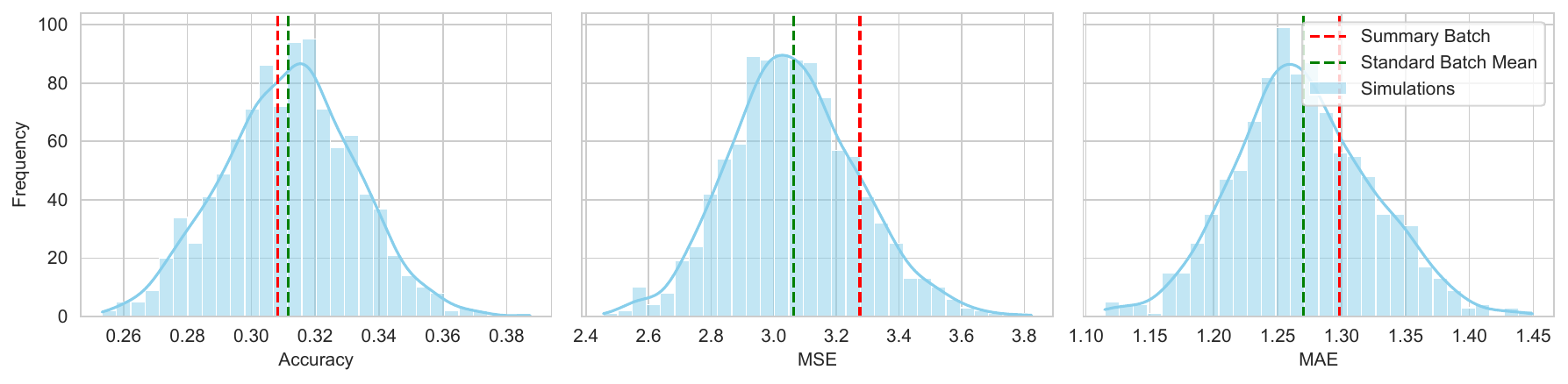}
    \caption{Distributions of Accuracy, Mean Squared Error (MSE), and Mean Absolute Error (MAE) across simulations.}
    \label{fig:simulation_metrics}
\end{figure*}

Previously, we used the \standard modality as the baseline, stacking judgments in the \summary modality to match the total time spent on assessments in the \standard batch. In this analysis, we reverse the approach by sampling judgments from the \standard modality to match the total time spent in the \summary batch, using the \summary modality as the baseline. We do that by using a simulation-based analysis. In each of the 1,000 simulations, we randomly sample assessments from the \standard modality to match the total time spent in the \summary modality. 
The results are shown in Figure~\ref{fig:simulation_metrics}. The average effectiveness scores achieved by the \standard modality across the simulations indicates that the \standard modality maintains a comparable level of accuracy even when constrained by the same amount of time. As in the previous case, no statistical significance emerges between the two modalities. These findings suggest that both modalities perform similarly in terms of absolute deviations from the ground truth, with no significant disparities in judgment quality.

\section{RQ3: Behavioral Insights}
\label{sect:rq-behavior}

\subsection{Usefulness and Need for Evidence}
\begin{figure*}[tb]
    \centering
    \begin{tabular}{c}
    \includegraphics[width=\linewidth]{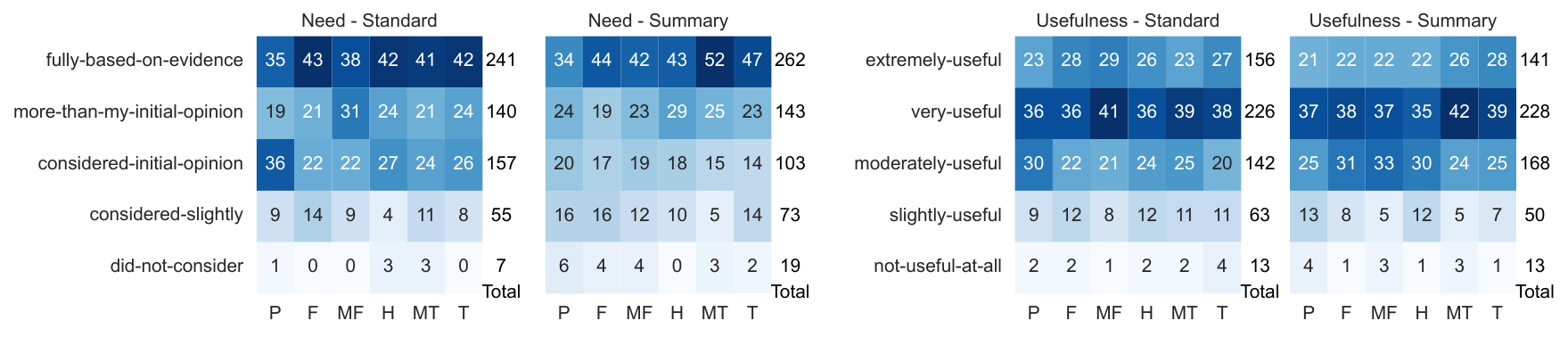}
    \end{tabular}
    \caption{Need and perceived usefulness of the evidence in the task.}
    \label{fig:need-usefulness}
\end{figure*}

In the context of crowdsourced truthfulness assessments, understanding the perceived usefulness of the evidence is important for understanding the quality and trustworthiness of the workers' provided assessments. This provides insight into how the evidence impacts the decision-making process and the reliance of participants on the provided information. 
Figure~\ref{fig:need-usefulness} shows the perceived usefulness and necessity of the evidence for the two modalities.

We start with inspecting the need for evidence as expressed by participants across different truthfulness levels. Data shows a significant reliance on evidence, with the majority of responses across both \standard and \summary modalities suggesting that assessments are ``fully based on evidence''. This strong dependency highlights the critical role that evidence plays in ensuring accurate and reliable assessments. Notably, in the summary modality, there is a slightly higher tendency for assessments to be fully based on evidence compared to the standard modality, suggesting that even summarized content sufficiently supports robust judgment formation. Less frequently, participants report that their assessments were ``more than my initial opinion'' or ``considered initial opinion'', indicating that while evidence guides their evaluations, preconceived notions or initial impressions also play a role. Interestingly, the number of participants who report their assessments as ``considered slightly'' or ``did not consider'' the evidence is minimal, suggesting the minimal influence of unsupported personal bias in the evaluation process.

Turning to the usefulness of evidence, participants across both modalities rate the majority of the evidence as being ``extremely useful'' or ``very useful''. This high level of perceived usefulness is consistent across all truthfulness categories, remarking the importance of quality evidence in the assessment task. In the \standard modality, there is a notable pattern where evidence is frequently considered at least ``moderately useful'', which aligns with the reduced amount of information presented to the workers, yet keeping its impact on their assessments. The less frequent ratings of ``slightly useful'' or ``not useful at all'' are significantly lower, which indicates that the evidence provided is overall useful to workers. This holds as well for the \summary modality: also summarized information appears to be adequately informative to workers.

Overall, these findings suggest the fundamental role of evidence in shaping the outcomes of crowdsourced truthfulness assessments. The high reliance on and usefulness of evidence in both \standard and \summary modalities suggest that effective summarization does not impact the value of the evidence presented to workers. 

\subsection{Interaction with Evidence}

We now detail the result of the inspection of the number of evidences participants interacted with in both the standard and summary modalities during the truthfulness assessment tasks.
We focus particularly on the number of evidences clicked by participants for each statement. A specific count of interest was 8, which indicates that participants clicked on exactly one link per statement (including gold questions), suggesting a uniform and minimal engagement across all statements provided in the task. For the \summary modality, we found that out of the total interactions, 511 instances (85.17\%) involved clicking exactly eight links to evidence. Similarly, in the \standard modality, there were 520 instances (86.67\%) where participants engaged with exactly eight links. This close similarity in percentages between the two modalities indicates a comparable level of engagement when workers are provided with either full or summarized content, in the case of workers who do the minimum amount of work. A similar pattern emerges when we exclude instances where participants clicked on exactly eight links. In such cases, the mean number of links clicked in the \summary modality was $18.70$, with a standard deviation of $8.09$, whereas in the \standard modality, the mean was slightly lower at $17.50$, with a narrower standard deviation of $3.39$. This is based on few (approx 15\%) of instances, and no statistical significance is measured.

We also extend the analysis to examine the maximum depth of links explored by participants, which reflects how deeply they looked into the provided evidences. We measured the maximum reached depth as the ordinal position of the farthest link clicked (e.g., clicking on the fifth link indicates a max depth of 5). It is again the case that the two modalities exhibit a similar behavior. In the \summary modality, 85\% of participants clicked no further than the first link, similar to 86.67\% in the \standard modality. When excluding these cases, participants in the \summary modality explored slightly more links on average (mean depth 3.63 vs. 3.43, no statistical significance detected), showing comparable engagement across both modalities.

\subsection{Interaction Across Evaluation Dimensions}

In addition to examining the depth of evidence exploration, we also analyzed the number of times each participant engaged with the assessment dimensions (i.e., truthfulness, usefulness of evidence, and need for evidence). Given that there are three dimensions in our evaluation setup, the minimum number of clicks required is three, one for each dimension. This measure helps us understand the decisiveness of participants and the extent to which they reconsider their assessments during the task.

Our findings show that in the \summary modality, 433 (72.17\%) of participants clicked each dimension only once, similar to 450 (75\%) in the \standard modality.
Excluding cases with exactly three clicks, the \summary modality shows a mean of $4.45$ clicks ($\sigma^2 = 0.67$), slightly lower than the \standard modality’s $4.60$ clicks ($\sigma^2 = 0.99$). No statistical significance is detected. These statistics suggest that participants in both modalities generally stick to their initial assessments, and the similar levels of engagement suggest that summarized content appears to be sufficient for making assessments. The slightly lower variance in the number of clicks in the \summary modality further reinforces this point, suggesting that participants are slightly more consistently confident in their assessments.

\begin{table}[tb]
\centering
    \caption{Correlation values between dimensions.}
    \label{tab:correlation_analysis}
    \begin{tabular}{lS[table-format=1.3]S[table-format=1.3]S[table-format=1.3]}
    \toprule
    \textbf{Modality} & \textbf{Need \& Usefulness } & \textbf{Need \& Time} & \textbf{Usefulness \& Time} \\
    \midrule
    \standard   & 0.483  & -0.020 & 0.030 \\
    \summary    & 0.481  & 0.004  & -0.097 \\
    \bottomrule
    \end{tabular}
\end{table}

To further explore the interplay between these dimensions, we conduct a correlation analysis examining the relationships among \textit{need}, \textit{usefulness}, and \textit{time elapsed} for both modalities (Table~\ref{tab:correlation_analysis}). The results reveal a consistent moderate positive correlation (\(\rho \approx 0.48\)) between \textit{need} and \textit{usefulness}, indicating that participants who perceive the evidence as more necessary also tend to consider it more useful, irrespective of whether they are presented with full-length or summarized content. This relationship underscores the complementary nature of these two dimensions, as both contribute to the assessment process. However, the moderate strength of the correlation suggests that while \textit{need} and \textit{usefulness} are related, they are not interchangeable. Instead, these dimensions capture distinct aspects of participants' interactions with evidence.
On the other hand, the correlations between \textit{need} and \textit{time elapsed}, and \textit{usefulness} and \textit{time elapsed}, are negligible (\(\rho \approx -0.02\) to \(-0.09\)), showing limited influence of these perceptions on task duration.

To more deeply analyze the interactions among the three assessment dimensions measured for the two modalities (i.e., \textit{time elapsed}, \textit{need}, and \textit{usefulness}), we performed an OLS regression analysis, augmented with \(\omega^2\) effect sizes \cite{olejnikAnova}, a well-established method in the literature for quantifying effect sizes \cite{ferro2016general,zampieri2019topic,ferro2018toward,roitero2020leveraging,ferro2019using}. The results confirm that the \summary modality significantly reduces task time when compared to the \standard modality (\(p < 0.05\), \(\omega^2 = 0.0036\)).
Conversely, perceptions of \textit{need}, \textit{usefulness}, and their interaction exhibited no statistically significant effects on task duration, with negligible \(\omega^2\) values (\(-0.0008\) for \textit{need}, \(-0.0001\) for \textit{usefulness}, and \(-0.0005\) for their interaction). 

\section{Conclusions and Future Work}
\label{sect:conclusion}

This study investigates the potential of summarized evidence to enhance the efficiency, effectiveness, and cost-effectiveness of crowdsourced truthfulness assessments.
For \textbf{RQ1 (Effectiveness)}, our results suggest that the \summary modality achieves accuracy and error metrics comparable to the \standard modality. While slight differences are observed in MSE and MAE, these are minimal and do not affect the practical utility of the \summary approach. These findings affirm that summarization effectively preserves the critical information necessary for accurate truthfulness assessments.

Regarding \textbf{RQ2 (Efficiency)}, we find that the \summary modality substantially reduces the time required for participants to complete their assessments compared to the \standard one. Workers using summarized evidence complete nearly double the number of judgments within the same time frame.
Addressing cost-effectiveness, we find that summarization significantly reduces the costs associated with large-scale truthfulness assessments. By enabling workers to complete more judgments within the same time frame, the \summary modality reduces overall task costs without reducing quality.

In terms of \textbf{RQ3 (Behavioral Insights)}, we observe that the \summary modality does not diminish participants' reliance on or perceived usefulness of evidence. Furthermore, the \summary modality produces higher internal agreement among workers, suggesting that the condensed presentation of information aligns participant judgments more effectively. This indicates that summarization not only maintains but also enhances certain aspects of the decision-making process, supporting consistent evidence-based assessments.

Despite these contributions, certain limitations must be acknowledged. First, the reliance on automatically generated summaries introduces the possibility of omitting critical nuances present in full-length evidence, which may impact worker judgment in complex cases. Additionally, our study focuses on a specific dataset and task context, which may limit the generalizability of our findings. Finally, while the \summary modality demonstrates promising efficiency, further research is needed to explore its long-term effects on worker fatigue, engagement, and the potential biases.

These findings also suggest directions for future work. We plan to refine summarization techniques to better tailor evidence to specific truthfulness categories or individual worker preferences; also, we plan to test the impact of evidence presentation on worker satisfaction and engagement. Additionally, we plan to generalize these methods to other domains, such as health misinformation or scientific claims. Finally, we plan to explore multi-document summarization techniques that would allow to present multiple pieces of evidence into a single, coherent summary.

\section*{Acknowledgments}

\includegraphics[width=1cm]{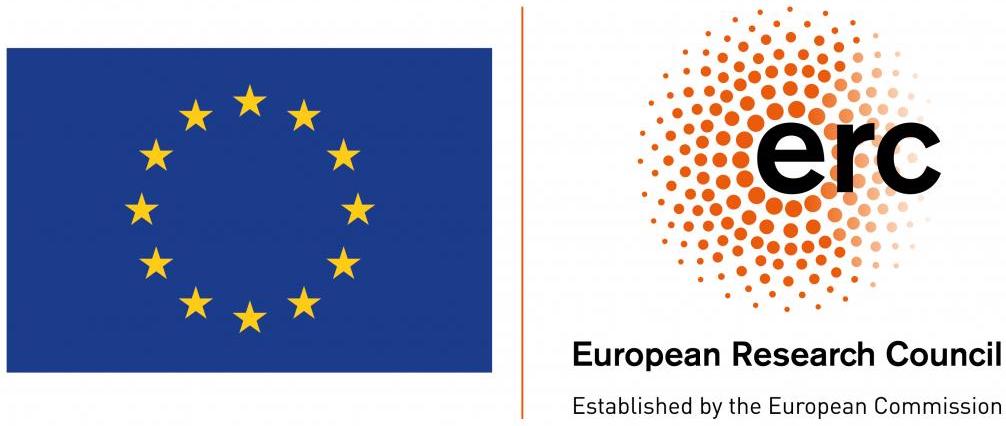} 
This research is partially funded by the European Union (ERC, ExplainYourself, 101077481), as well as a Danish Data Science Academy postdoctoral fellowship (grant number: 2023-1425) and is supported by the Pioneer Centre for AI, DNRF grant number P1.
It is also partially supported by the PRIN 2022 Project, ``MoT—The Measure of Truth: An Evaluation-Centered Machine-Human Hybrid Framework for Assessing Information Truthfulness'', Code No. 20227F2ZN3, CUP No. G53D23002800006, funded by the European Union – Next Generation EU – PNRR M4 C2 I1.1.

\clearpage
\bibliographystyle{ACM-Reference-Format}
\bibliography{main}

\end{document}